# Dual-polarization RF Channelizer Based on Kerr Soliton Microcomb Sources

Xingyuan Xu, David J. Moss

*Abstract*—We report a dual-polarization radio frequency (RF) channelizer based on microcombs. With the tailored mismatch between the FSRs of the active and passive MRRs, wideband RF spectra can be channelized into multiple segments featuring digital-compatible bandwidths via the Vernier effect. Due to the use of dual-polarization states, the number of channelized spectral segments, and thus the RF instantaneous bandwidth (with a certain spectral resolution), can be doubled. In our experiments, we used 20 microcomb lines with ~ 49 GHz FSR to achieve 20 channels for each polarization, with high RF spectra slicing resolutions at 144 MHz (TE) and 163 MHz (TM), respectively; achieving an instantaneous RF operation bandwidth of 3.1 GHz (TE) and 2.2 GHz (TM). Our approach paves the path towards monolithically integrated photonic RF receivers (the key components—active and passive MRRs are all fabricated on the same platform) with reduced complexity, size, and unprecedented performance, which is important for wide RF applications with digital-compatible signal detection.

*Index Terms*—Microwave photonic, signal channelization, integrated optical frequency comb.

## I. INTRODUCTION

CHANNELIZED receivers are key building blocks of modern radio frequency (RF) systems including modern electronic warfare systems, deep space tracking and telecommunications [1-4], as they enable the detection of wideband RF spectra with digital-compatible devices that have much lower bandwidths (such as generic analog-to-digital converters at hundreds of MHz). However, multi-format, multi-frequency and wideband signals pose challenges to present electronic processing systems. While electronic RF channelizers (generally formed by filter banks) are subject to the bandwidth bottleneck, photonic approaches are promising since they can offer ultra-large bandwidths, low transmission loss and strong immunity to electromagnetic interference.

Photonic RF channelizers can generally be divided into three categories. The first category depends on the spectral-to-spatial conversion performed by a diffraction grating or an integrated Fresnel lens, yielding multiple parallel channels containing different frequency components in free space [5, 6]. This method, while advantageous in terms of instantaneous bandwidth, poses limitations in the system's overall footprint and resolution. The second category is more compact, which employs the Vernier effect between a multi-wavelength source and a periodic filter array to achieve high-resolution wideband RF spectral channelization. Several types of devices and platforms have been utilized for this purpose, such as discrete laser arrays [7], parametric processes in nonlinear fiber [8], cascaded electro-optic modulators [9] and so on. However, the above approaches are restricted in channel number, spectral resolution, and compatibility for monolithic integration. The last solution is based on a frequency-to-time mapping, where the RF spectra are mapped to the time domain by using wavelength scanning or frequency shifting [10]. Each wavelength is marked with a specific channel at the corresponding time slot, and only one photodetector (PD) is required to obtain the RF signal at a fast scan rate. Yet those approaches face limitations in one form or another — the scanning frequency step determines final channel number and operating bandwidth, even the serial detected results are non-contiguous.

Recently, microcombs, especially CMOS-compatible microcombs [11], have shown unique advantages over conventional mode-locked fiber combs [12] and electro-optical combs [9] as they provide massively coherent wavelength channels at the chip-scale size and have proven to be widely used in microwave photonics [13-15].

X. Xu, is with the State Key Laboratory of Information of Photonics and Optical Communications, School of Electronic Engineering, Beijing University of Posts and Telecommunications, Beijing 100876, China.

D. J. Moss is with Optical Sciences Centre, Swinburne University of Technology, Hawthorn, VIC 3122, Australia (e-mail: dmoss@swin.edu.au).

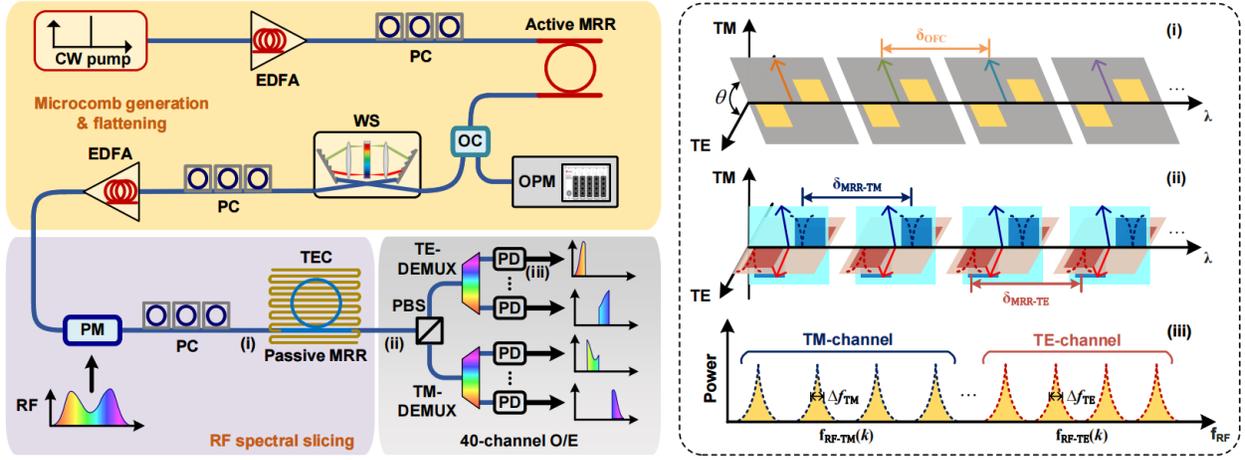

Fig. 1. Schematic diagram of 40-channel dual-polarization RF channelizer based on microcomb. EDFA: erbium-doped fibre amplifier. PC: polarization controller. MRR: micro-ring resonator. OC: optical coupler. OPM: optical powermeter. WS: WaveShaper. PM: phase modulator. TEC: temperature controller. PBS: polarization beam splitter. DEMUX: demultiplexer PD: photodetector.

Here, we first leverage the polarization division of integrated photonics for RF channelization, and report a dual-polarization photonic RF channelizer, achieved with two MRRs with slightly different FSRs at ~49 GHz. The first MRR is used to generate optical comb lines (over 80 in the C band and 20 were used in this work), while the second acts as dual narrowband notch filters in two polarization states to slice the RF spectrum. The high Q passive MRR offers narrow resonance linewidths of 144 MHz (TE) and 163 MHz (TM), enabling high-resolution RF channelization and thus lowered requirements of subsequent analog-to-digital converters (ADCs) for digital processing; in addition, with tailored FSR mismatch between the active and passive MRRs, the RF channelization steps (~163 MHz for TE and ~117 MHz for TM) are closer to the slicing resolution, leading to an improved channel crosstalk of ~12 dB. Most importantly, the use of dual polarization modes doubled the number of channels (40 in total, 20 for each polarization state) and instantaneous bandwidth (3.1 GHz for TE and 2.2 GHz for TM in this work) in contrast to those using a single polarization mode, with wideband operation verified via thermal tuning of the passive MRR. This approach explores the polarization division of optics using integrated devices, further demonstrating the potentials of photonic channelizers for wideband RF signal processing.

## II. PRINCIPLE

Figure 1 depicts the schematic of the 40-channel dual-polarization RF channelizer that consists of three modules. The first module achieves microcomb generation and flattening, where an active MRR is pumped by a continuous-wave (CW) laser and amplified by an erbium-doped fiber amplifier (EDFA) to excite intracavity parametric oscillations. The MRR features high Q-factor, high nonlinear coefficients and designed anomalous dispersion, providing sufficient parametric gain to generate a Kerr frequency comb. Here, we utilize the soliton crystal combs and perform spectral shaping by a commercial WaveShaper to achieve equalized channel power.

Assuming the pre-shaped lines are generated with a spacing of $\delta_{\text{OFC}}$, the optical frequency of the $k_{th}$ ($k = 1, 2, 3, …, 20$) comb

line can be written as

$$f_{\text{OFC}}(k)=f_{\text{OFC}}(1)+(k-1)\delta_{\text{OFC}} \tag{1}$$

where $f_{\text{OFC}}(1)$ is the frequency of the first comb line.

In the second module, the flattened combs are fed into an electro-optic phase modulator, where the broadband RF spectra are multicast onto each wavelength channel. Next, the copied RF signals are sliced into segments by a dual-polarization passive MRR with FSRs of $\delta_{\text{MRR-TE}}$ and $\delta_{\text{MRR-TM}}$ for TE and TM polarization, respectively, where the channel resolution is determined by the 3 dB bandwidth of the TE- and TM-polarization resonances, denoted as $\Delta f_{\text{TE}}$ and $\Delta f_{\text{TM}}$.

The $k_{th}$ centre frequency of the passive MRR's resonance follows

$$f_{\text{MRR-TE}}(k)=f_{\text{MRR-TE}}(1)+(k-1)\delta_{\text{MRR-TE}} \tag{2}$$

$$f_{\text{MRR-TM}}(k)=f_{\text{MRR-TM}}(1)+(k-1)\delta_{\text{MRR-TM}} \tag{3}$$

where $f_{\text{MRR-TE}}(1)$ and $f_{\text{MRR-TM}}(1)$ are the frequency of the first filtering transmission lines of two polarizations, $\delta_{\text{MRR-TE}}$ and $\delta_{\text{MRR-TM}}$ denote the FSRs of passive MRR at TE- and TM-polarization.

The detailed mechanism of the dual polarization RF photonic channelizer is described in right part of Figure 1. For each channel, the microcomb is phase-modulated by RF signals to produce counter-phase sidebands whose offset angle from the TE polarization is marked as $\theta$ (Fig. 1(i)). The components of modulated signal falling in TE and TM-planes are filtered by the orthogonally-polarized notch filters (i.e., the TE- and TM-resonances of the passive MRR), respectively (Fig. 1(ii)), which break the balance between the inverted sidebands and enable the phase-to-intensity modulation conversion. Finally, the dual polarization modes are separated by a polarization beam splitter (PBS), followed by wavelength-division demultiplexers to separate the wavelength channels for photodetection. After parallel detection, the channelized RF signals are each centered at $f_{\text{RF-TE}}(k)$ (or $f_{\text{RF-TM}}(k)$) with a spectral bandwidth of $\Delta f_{\text{TE}}$ (or $\Delta f_{\text{TM}}$), within the operation bandwidth of generic ADCs (Fig. 1(iii)).

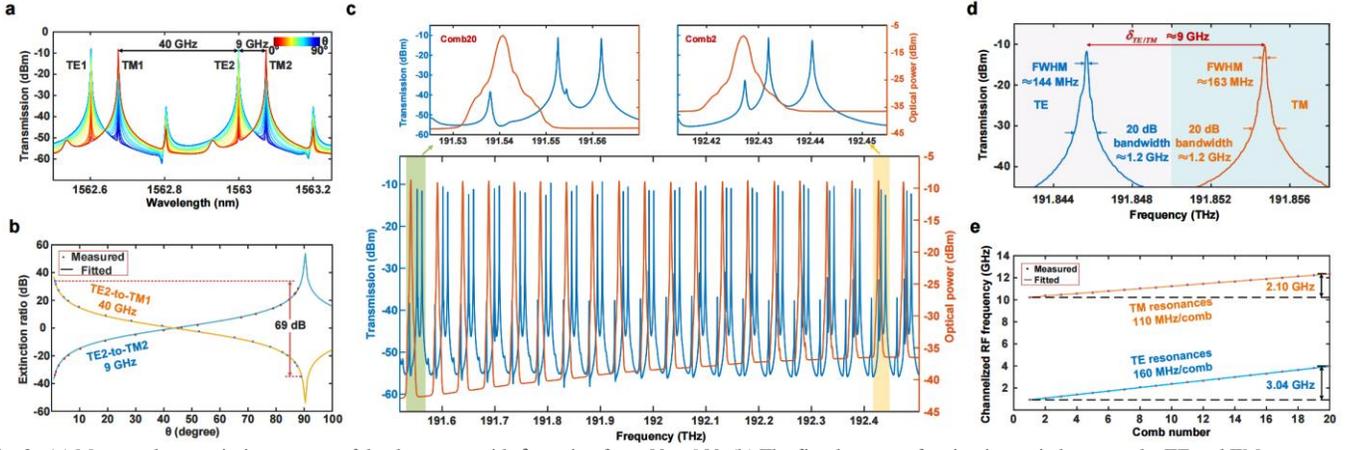

Fig. 2. (a) Measured transmission spectra of the drop-port with $\theta$ varying from 0° to 90°. (b) The fitted curves of extinction ratio between the TE and TM resonances as $\theta$ varies. (c) The experimental 40-channel channelizer of 20 microcombs optical spectrum (orange lines) and drop-port dual-polarization transmission spectrum (blue lines) of passive MRR. The zoom-in views of the shaded areas indicate the relative frequency spacing between the comb and adjacent transmission resonance. (d) Drop-port transmission spectrum of the active MRR showing TE and TM resonances with FWHM of 144 MHz and 163 MHz, respectively, corresponding to Q factors over $1.2 \times 10^6$. (e) The fitted channelization slopes are 160 MHz (TE) and 110 MHz (TM) per comb wavelength.

Therefore, the progressive RF centre frequencies of the RF spectra on the TE- and TM-channel, respectively, are given by

$$f_{RF\text{-}TE}(k) = f_{MRR\text{-}TE}(k) - f_{OFC}(k)$$
$$= [f_{MRR\text{-}TE}(1) - f_{OFC}(1)] + (k-1)(\delta_{MRR\text{-}TE} - \delta_{OFC}) \quad (4)$$

$$f_{RF\text{-}TM}(k) = f_{MRR\text{-}TM}(k) - f_{OFC}(k)$$
$$= [f_{MRR\text{-}TM}(1) - f_{OFC}(1)] + (k-1)(\delta_{MRR\text{-}TM} - \delta_{OFC}) \quad (5)$$

where $f_{RF\text{-}TE}(k)$ and $f_{RF\text{-}TM}(k)$ are the $k_{th}$ channelized RF centre frequencies of TE- and TM-channel, $[f_{MRR\text{-}TE}(1) - f_{OFC}(1)]$ and $[f_{MRR\text{-}TM}(1) - f_{OFC}(1)]$ denote the relative spacing between the first comb and the adjacent dual polarization resonances, namely, the offset of the channelized RF frequency. And $(\delta_{MRR\text{-}TE} - \delta_{OFC})$ and $(\delta_{MRR\text{-}TM} - \delta_{OFC})$ represent the channelized RF frequency step between adjacent wavelength channels for TE- and TM-channel, respectively.

We further analyzed the operation of the dual-polarization passive MRR using the Jones matrix [16], and the through-port transmission can be written by

$$R = \begin{bmatrix} T_{TE} & 0 \\ 0 & T_{TM} \end{bmatrix} \quad (6)$$

where $T_{TE}$ and $T_{TM}$ are the through-port transfer functions of the passive MRR given by

$$T_{TE} = \frac{t(1 - ae^{i\phi_{TE}})}{1 - t^2 ae^{i\phi_{TE}}} \quad (7)$$

$$T_{TM} = \frac{t(1 - ae^{i\phi_{TM}})}{1 - t^2 ae^{i\phi_{TM}}} \quad (8)$$

where $t$ is the transmission coefficient between the bus waveguide and the passive MRR, $a$ is the round-trip transmission factor, $\phi_{TE} = 2\pi L \times n_{eff\_TE} / \lambda$ and $\phi_{TM} = 2\pi L \times n_{eff\_TM} / \lambda$ are the single-pass phase shifts of TE and TM modes, respectively, $L$ denotes the round-trip length, $n_{eff\_TE}$ and $n_{eff\_TM}$ denote the effective indices of TE and TM modes, and $\lambda$ represents the wavelength.

The phase-modulated optical signal can be given as

$$E_0 \begin{bmatrix} \cos\theta \\ \sin\theta \end{bmatrix} \quad (9)$$

where $E_0$ denotes the modulated optical signal, $\theta$ is the polarization angle relative to the TE-axis, so the output field of the passive MRR can be written as

$$E_{out} = RE_0 \begin{bmatrix} \cos\theta \\ \sin\theta \end{bmatrix} = E_0 \begin{bmatrix} T_{TE} \cdot \cos\theta \\ T_{TM} \cdot \sin\theta \end{bmatrix} \quad (10)$$

According to the above equation, the optical power of TE- and TM-polarized optical signals are proportional to $\cos^2\theta$ and $\sin^2\theta$, respectively. Hence the extinction ratio between the channelized RF spectral segments of TE- and TM-resonances is given by

$$ER(\theta) \propto \cot^2\theta \quad (11)$$

$ER(\theta)$ can be continuously adjusted by changing $\theta$, limited only by the performance of the polarization controller. Moreover, $\cot^2\theta$ can infinitely approach 0 or infinity as $\theta$ approaches 90° or 0°, thus theoretically an ultra-large dynamic tuning range of the extinction ratio for dual polarization states can be expected. Specifically, when $\theta = 45°$, the amplitude of TE and TM polarization is equal.

## III. EXPERIMENTAL RESULTS

Both active and passive MRRs are fabricated on a high-index doped silica glass platform through a CMOS compatible fabrication process [17]. The two MRRs featured similar characteristics with radii of ~592 μm, corresponding to a FSR of ~0.4 nm (~49 GHz), with Q factors over one million. To demonstrate the analysis above regarding the dual-polarization passive MRR, we measured its transmission spectra with different polarization angles $\theta$. As $\theta$ varies from 0° to 90°, the transmission spectra evolution of the passive MRR (TE and TM resonances) and extinction ratios between the orthogonally polarized resonances are shown in Fig. 2(a-b). We note that the polarization angle $\theta$ is deferred by comparing measured and simulated results. As for the extinction ratio of dual modes (TE2-to-TM1 and TE2-to-TM2), the former varies from 34 dB to -35 dB, which indicates a continuously tunable extinction ratio of over 69 dB. Note that we actually used the through-port transmission of the passive MRR to achieve the channelizer, although here we plotted the drop-port transmission spectra to reveal the relationships between $\theta$ and the extinction ratio between the orthogonally polarized resonances.

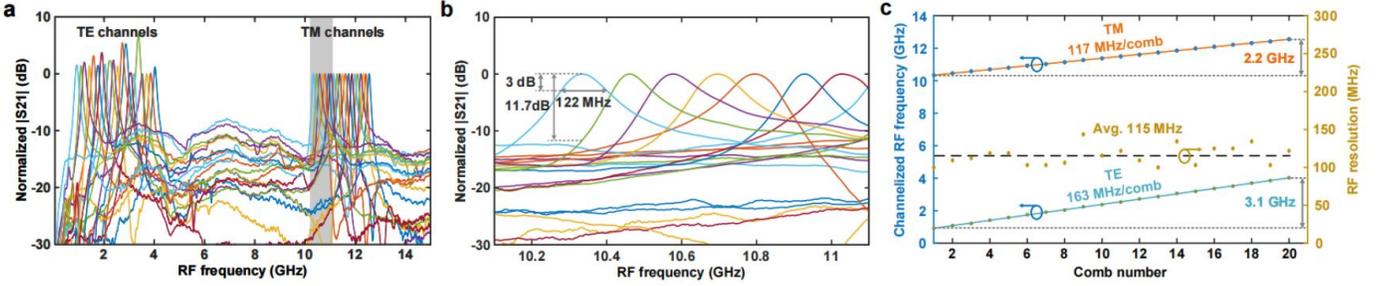

Fig. 3. Experimental RF transmission spectra of (a) the dual-polarization 40 channels and (b) a zoom-in view of the shaded area in (a). The first channel of TM mode with a 3 dB bandwidth of 122 MHz, and the adjacent channel crosstalk over 10 dB. (c) Derived channelized RF frequency of TE and TM passbands and RF resolution of the channelizer.

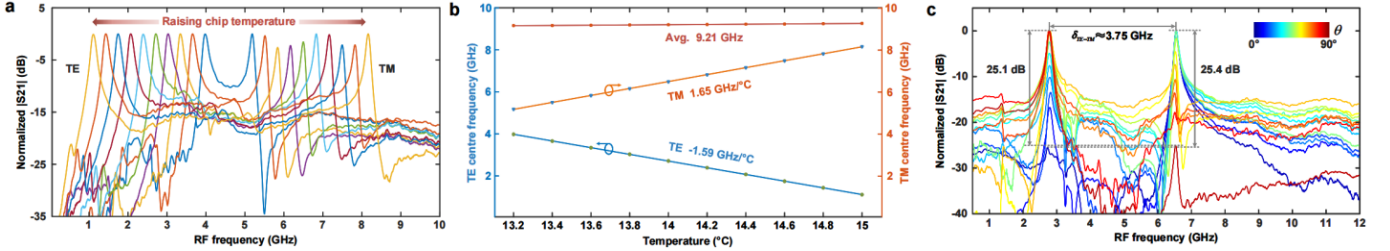

Fig. 4. (a) Experimental RF transmission spectra of a dual-polarization channel with adjusting the passive chip temperature. (b) Derived centre frequencies of TE- and TM-channel with varying temperature. (c) Experimental RF transmission spectra with varying polarization angle $\theta$ between TE- and TM-channel, resulting in varying extinction ratio.

The measured drop-port transmission spectra at dual polarization of the passive MRR and 20 microcomb lines are shown in Fig. 2(c). The microcombs are pre-shaped by an optical programmable processor. Proper polarization state allows both TE and TM mode resonances to be observed simultaneously, as the blue lines mark, while the orange lines denote the flattened combs. The zoom-in views show the details near the second and the 20th comb lines. We note that the combs' linewidths are much smaller (potentially at kHz level) than those shown in the figure (limited by the resolution of the optical spectrum analyzer at 0.02 nm). As Fig. 2(d) depicts, the full-width at half-maximum (FWHM) of TE and TM resonance are ~144 MHz and ~163 MHz, respectively, corresponding to a high Q factor over 1.2 million, and a 20 dB bandwidth of ~ 1.2 GHz, which leads a high RF spectral resolution for the channelizer and thus reduced bandwidth requirements to ADCs (<200 MHz).

The channelized RF frequencies $f_{RF}(k)$ of dual polarization are derived, corresponding to the spacing between the comb and adjacent TE and TM resonances, showing an upward trend from red to blue in Fig. 2(e). Due to the insufficient resolution of the spectrometer, here the peak wavelength of each comb is calculated using the inherent FSR of the MRR. The fitted results indicate that the channelized RF frequencies increase at 160 MHz (TE) and 110 MHz (TM) per channel, eventually achieving instantaneous operating bandwidths of 3.04 GHz (TE) and 2.1 GHz (TM), 5.14 GHz in total. We note that a wider instantaneous bandwidth can be achieved with more wavelength channels, for example, with 80 channels the instantaneous bandwidths can be increased by 4 times to over 20 GHz, sufficient for general RF applications.

Next, the frequency responses of the 40 channels under dual polarization are verified by VNA in the RF domain, as shown in Figure 3. The 3dB bandwidth of RF channels, or the achieved RF channelizing resolution, are ~ 115 MHz. We note that due to imperfections of devices across a broad optical bandwidth, the measured RF channels can have power fluctuations, although they can be equalized straightforwardly by adjust the optical power of each comb line during the flattening process. As shown in Fig. 3(c), the channelized RF frequencies (i.e., the center frequencies of each RF channel) indicate instantaneous operation bandwidths of 3.1 GHz and 2.2 GHz for TE and TM polarization, respectively, with RF channelization steps of 163 MHz (TE) and 117 MHz (TM), which closely matches with the results in Fig. 2(e). We note that, due to the relatively close match between the channelized RF frequencies' step and the resolution, the crosstalk between adjacent RF channels is further reduced (~12 dB, as shown in Fig. 3(b)), in contrast to our previous work [15]. To further reduce the crosstalk, several solutions can be adopted [15]: (i) tailored passive MRR's FSR via accurate design and nanofabrication allows matching between the RF channelization step and resolution; (ii) passive optical filters with higher roll-offs and flat passbands, which can be achieved by high-order cascaded MRRs [18].

To further verify the tunability of the proposed RF channelizer for spectral analysis at diverse RF bands (such as frequency-upconverted baseband RF signals), we thermally tuned the passive MRR to change the offset or spectral interval between a comb line and its adjacent passive resonances. Figure 4 shows the measured RF transmission spectra. The thermal tuning efficiencies of TE and TM channels are -1.59 GHz/°C and 1.65 GHz/°C, respectively, shown as in Fig. 4(b), indicating a wide tuning range up to >60 GHz with ~40 °C temperature variation, well within the capability of external or on-chip heaters and sufficient for wideband RF applications. Furthermore, to verify the flexibility in the extinction ratio between TE- and TM-polarized channels, which can be used to equalize the channels' power, we continuously adjusted the polarization angle $\theta$ via a polarization controller. By varying from 0 to 90 degree, the extinction ratio between the TE- and TM- RF channels varies from -25.1 to 25.4 dB, as shown in Fig. 4(c). This result indicates that the TE- and TM- RF channels can be equalized or switched on/off, bringing additional flexibilities for post sub-bands receiving and processing.

We note that the instantaneous RF operation bandwidth is given by the product of channel number and the channelizing resolution, as shown in Fig. 2(e) and Fig. 3(c), thus it can be further increased by: a) increasing the number of channels within available optical bands — this can be achieved by using active/passive MRRs with wider optical bandwidths and smaller FSRs, although we note that smaller FSRs introduces tradeoffs with the Nyquist bandwidth (i.e., half of the MRR's FSR); b) decreasing the channelizing resolution — this can be achieved by using passive MRRs with lower Q factor (ideally high-order MRRs with wider passbands), although this imposes higher requirements on the performance of ADCs, thus needs to be chose upon the practical systems' requirements of the instantaneous bandwidth, cost and size; c) the optical spectral interval between adjacent resonances of the passive MRR/the FSR of the comb source — this can be enlarged by using MRRs with higher FSRs [19], nonetheless, this brings about tradeoffs that larger FSRs lead to less channels within a certain optical bandwidth. These devices will benefit from the advances made in microcombs generally and microwave applications of microcombs. [20-82]

## IV. CONCLUSION

In this work, we report a dual-polarization photonic RF channelizer using two MRRs with slightly different FSRs. The first MRR is used to generate optical comb lines, while the second acts as dual narrowband notch filters in two polarizations to slice the RF spectrum. We achieved high RF channelization resolution of 144 MHz (TE), and 163 MHz (TM), and doubled instantaneous bandwidth (3.1 GHz for TE and 2.2 GHz for TM) due to the use of dual polarization states. The tunability of the proposed RF channelizer in terms of the operation bandwidth and extinction ratio between TE- and TM- channels are also experimentally verified, offering additional flexibilities for tailored RF systems. This approach explores the polarization division of integrated devices for photonic RF channelizers, which has the full potential to be monolithically integrated with ADCs and enables unprecedented performances of RF systems.

## REFERENCES

[1] G. W. Anderson, D. C. Webb, A. E. Spezio, and J. N. Lee, "Advanced channelization for RF, microwave, and millimeterwave applications," *Proc. IEEE*, vol. 79, no. 3, pp. 355–388, Mar. 1991.

[2] J. Capmany and D. Novak, "Microwave photonics combines two worlds," *Nature Photon*, vol. 1, no. 6, pp. 319–330, Jun. 2007.

[3] J. Yao, "Microwave Photonics," *J. Lightwave Technol.*, vol. 27, no. 3, pp. 314–335, Feb. 2009.

[4] X. Zou, B. Lu, W. Pan, L. Yan, A. Stöhr, and J. Yao, "Photonics for microwave measurements: Photonics for microwave measurements," *Laser & Photonics Reviews*, vol. 10, no. 5, pp. 711–734, Sep. 2016.

[5] Wenshen Wang, R. L. Davis, T. J. Jung, R. Lodenkamper, L. J. Lembo, J. C. Brock, and M. C. Wu, "Characterization of a coherent optical RF channelizer based on a diffraction grating," *IEEE Trans. Microwave Theory Techn.*, vol. 49, no. 10, pp. 1996–2001, Oct. 2001.

[6] S. T. Winnall, A. C. Lindsay, M. W. Austin, J. Canning, and A. Mitchell, "A microwave channelizer and spectroscope based on an integrated optical Bragg-grating Fabry-Perot and integrated hybrid Fresnel lens system," *IEEE Trans. Microwave Theory Techn.*, vol. 54, no. 2, pp. 868–872, Feb. 2006.

[7] S. J. Strutz and K. J. Williams, "An 8-18-GHz all-optical microwave downconverter with channelization," *IEEE Trans. Microwave Theory Techn.*, vol. 49, no. 10, pp. 1992–1995, Oct. 2001.

[8] C.-S. Brès, S. Zlatanovic, A. O. J. Wiberg, and S. Radic, "Reconfigurable parametric channelized receiver for instantaneous spectral analysis," *Opt. Express*, vol. 19, no. 4, p. 3531, Feb. 2011.

[9] Xiaojun Xie, Yitang Dai, Kun Xu, Jian Niu, Ruixin Wang, Li Yan, and Jintong Lin, "Broadband Photonic RF Channelization Based on Coherent Optical Frequency Combs and I/Q Demodulators," *IEEE Photonics J.*, vol. 4, no. 4, pp. 1196–1202, Aug. 2012.

[10] R. Li, H. Chen, Y. Yu, M. Chen, S. Yang, and S. Xie, "Multiple-frequency measurement based on serial photonic channelization using optical wavelength scanning," *Opt. Lett.*, vol. 38, no. 22, p. 4781, Nov. 2013.

[11] P. Del'Haye, A. Schliesser, O. Arcizet, T. Wilken, R. Holzwarth, and T. J. Kippenberg, "Optical frequency comb generation from a monolithic microresonator," *Nature*, vol. 450, no. 7173, pp. 1214–1217, Dec. 2007.

[12] N. R. Newbury and W. C. Swann, "Low-noise fiber-laser frequency combs (Invited)," *J. Opt. Soc. Am. B*, vol. 24, no. 8, p. 1756, Aug. 2007.

[13] X. Xu, J. Wu, T. G. Nguyen, T. Moein, S. T. Chu, B. E. Little, R. Morandotti, A. Mitchell, and D. J. Moss, "Photonic microwave true time delays for phased array antennas using a 49 GHz FSR integrated optical micro-comb source [Invited]," *Photonics Research*, vol. 6, no. 5, pp. B30-B36, May 1. 2018.

[14] X. Xu, J. Wu, T. G. Nguyen, S. T. Chu, B. E. Little, R. Morandotti, A. Mitchell, and D. J. Moss, "Broadband RF Channelizer Based on an Integrated Optical Frequency Kerr Comb Source," *J. Lightwave Technol.*, vol. 36, no. 19, pp. 4519–4526, Oct. 2018.

[15] X. Xu, M. Tan, J. Wu, A. Boes, T. G. Nguyen, S. T. Chu, B. E. Little, R. Morandotti, A. Mitchell, and D. J. Moss, "Broadband Photonic RF Channelizer With 92 Channels Based on a Soliton Crystal Microcomb," *J. Lightwave Technol.*, vol. 38, no. 18, pp. 5116–5121, Sep. 2020.

[16] P. Bianucci, C. R. Fietz, J. W. Robertson, G. Shvets, and C.-K. Shih, "Whispering gallery mode microresonators as polarization converters," *Opt. Lett.*, vol. 32, no. 15, p. 2224, Aug. 2007.

[17] D. J. Moss, R. Morandotti, A. L. Gaeta, and M. Lipson, "New CMOS-compatible platforms based on silicon nitride and Hydex for nonlinear optics," *Nature Photon*, vol. 7, no. 8, pp. 597–607, Aug. 2013.

[18] B. E. Little et al., "Very high-order microring resonator filters for WDM applications," IEEE Photon. Technol. Lett., vol. 16, no. 10, pp. 2263–2265, Oct. 2004.

[19] J. S. Levy, A. Gondarenko, M. A. Foster, A. C. Turner-Foster, A. L. Gaeta, and M. Lipson, "CMOS-compatible multiple-wavelength oscillator for on-chip optical interconnects," *Nat. Photonics*, vol. 4, no. 1, pp. 37-40, Jan. 2010.

20. M.Ferrera et al., "CMOS compatible integrated all-optical RF spectrum analyzer", Optics Express, vol. 22, no. 18, 21488 - 21498 (2014).

21. M. Kues, et al., "Passively modelocked laser with an ultra-narrow spectral width", Nature Photonics, vol. 11, no. 3, pp. 159, 2017.

22. L. Razzari, et al., "CMOS-compatible integrated optical hyper-parametric oscillator," Nature Photonics, vol. 4, no. 1, pp. 41-45, 2010.

23. M. Ferrera, et al., "Low-power continuous-wave nonlinear optics in doped silica glass integrated waveguide structures," Nature Photonics, vol. 2, no. 12, pp. 737-740, 2008.

24. M.Ferrera et al."On-Chip ultra-fast 1st and 2nd order CMOS compatible all-optical integration", Opt. Express, vol. 19, (23)pp. 23153-23161 (2011).

25. D. Duchesne, M. Peccianti, M. R. E. Lamont, et al., "Supercontinuum generation in a high index doped silica glass spiral waveguide," Optics Express, vol. 18, no, 2, pp. 923-930, 2010.

26. H Bao, L Olivieri, M Rowley, ST Chu, BE Little, R Morandotti, DJ Moss, ... "Turing patterns in a fiber laser with a nested microresonator: Robust and controllable microcomb generation", Physical Review Research **2** (2), 023395 (2020).

27. M. Ferrera, et al., "On-chip CMOS-compatible all-optical integrator", Nature Communications, vol. 1, Article 29, 2010.

28. A. Pasquazi, et al., "All-optical wavelength conversion in an integrated ring resonator," Optics Express, vol. 18, no. 4, pp. 3858-3863, 2010.

29. A.Pasquazi, Y. Park, J. Azana, et al., "Efficient wavelength conversion and net parametric gain via Four Wave Mixing in a high index doped silica waveguide," Optics Express, vol. 18, no. 8, pp. 7634-7641, 2010.

30. M. Peccianti, M. Ferrera, L. Razzari, et al., "Subpicosecond optical pulse compression via an integrated nonlinear chirper," Optics Express, vol. 18, no. 8, pp. 7625-7633, 2010.

31. Little, B. E. et al., "Very high-order microring resonator filters for WDM applications", IEEE Photonics Technol. Lett. **16**, 2263–2265 (2004).

32. M. Ferrera et al., "Low Power CW Parametric Mixing in a Low Dispersion High Index Doped Silica Glass Micro-Ring Resonator with Q-factor > 1 Million", Optics Express, vol.17, no. 16, pp. 14098–14103 (2009).

33. M. Peccianti, et al., "Demonstration of an ultrafast nonlinear microcavity modelocked laser", Nature Communications, vol. 3, pp. 765, 2012.

34. A.Pasquazi, et al., "Self-locked optical parametric oscillation in a CMOS compatible microring resonator: a route to robust optical frequency comb generation on a chip," Optics Express, vol. 21, no. 11, pp. 13333-13341, 2013.

35. A.Pasquazi, et al., "Stable, dual mode, high repetition rate mode-locked laser based on a microring resonator," Optics Express, vol. 20, no. 24, pp. 27355-27362, 2012.

36. Pasquazi, A. et al. Micro-combs: a novel generation of optical sources. Physics Reports **729**, 1-81 (2018).

37. H. Bao, et al, Laser cavity-soliton microcombs, Nature Photonics, vol. 13, no. 6, pp. 384-389, Jun. 2019.

38. Antonio Cutrona, Maxwell Rowley, Debayan Das, Luana Olivieri, Luke Peters, Sai T. Chu, Brent L. Little, Roberto Morandotti, David J. Moss, Juan Sebastian Totero Gongora, Marco Peccianti, Alessia Pasquazi, "High Conversion Efficiency in Laser Cavity-Soliton Microcombs", Optics Express Vol. 30, Issue 22, pp. 39816-39825 (2022). https://doi.org/10.1364/OE.470376.

39. M.Rowley, P.Hanzard, A.Cutrona, H.Bao, S.Chu, B.Little, R.Morandotti, D. J. Moss, G. Oppo, J. Gongora, M. Peccianti and A. Pasquazi, "Self-emergence of robust solitons in a micro-cavity", Nature **608** (7922) 303–309 (2022).

40. A. Cutrona, M. Rowley, A. Bendahmane, V. Cecconi,L. Peters, L. Olivieri, B. E. Little, S. T. Chu, S. Stivala, R. Morandotti, D. J. Moss, J. S. Totero-Gongora, M. Peccianti, A. Pasquazi, "Nonlocal bonding of a soliton and a blue-detuned state in a microcomb laser", *Nature Communications Physics* **6** (2023).

41. A. Cutrona, M. Rowley, A. Bendahmane, V. Cecconi,L. Peters, L. Olivieri, B. E. Little, S. T. Chu, S. Stivala, R. Morandotti, D.J Moss, J. S. Totero-Gongora, M. Peccianti, A. Pasquazi, "Stability Properties of Laser Cavity-Solitons for Metrological Applications", *Applied Physics Letters* **122** (12) 121104 (2023); doi: 10.1063/5.0134147.X. Xu, J. Wu, M. Shoeiby, T. G. Nguyen, S. T. Chu, B. E. Little, R. Morandotti, A. Mitchell, and D. J. Moss, "Reconfigurable broadband microwave photonic intensity differentiator based on an integrated optical frequency comb source", *APL Photonics*, vol. 2, no. 9, 096104 , Sep. 2017.

42. Xu, X., et al., Photonic microwave true time delays for phased array antennas using a 49 GHz FSR integrated micro-comb source, *Photonics Research*, **6**, B30-B36 (2018).

43. X. Xu, M. Tan, J. Wu, R. Morandotti, A. Mitchell, and D. J. Moss, "Microcomb-based photonic RF signal processing", *IEEE Photonics Technology Letters*, vol. 31 no. 23 1854-1857, 2019.

44. Xu, et al., "Advanced adaptive photonic RF filters with 80 taps based on an integrated optical micro-comb source," *Journal of Lightwave Technology*, vol. 37, no. 4, pp. 1288-1295 (2019).

45. X. Xu, et al., "Photonic RF and microwave integrator with soliton crystal microcombs", *IEEE Transactions on Circuits and Systems II: Express Briefs*, vol. 67, no. 12, pp. 3582-3586, 2020. DOI:10.1109/TCSII.2020.2995682.

46. X. Xu, et al., "High performance RF filters via bandwidth scaling with Kerr micro-combs", *APL Photonics*, vol. 4 (2) 026102. 2019.

47. M. Tan, et al., "Microwave and RF photonic fractional Hilbert transformer based on a 50 GHz Kerr micro-comb", *Journal of Lightwave Technology*, vol. 37, no. 24, pp. 6097 – 6104, 2019.

48. M. Tan, et al., "RF and microwave fractional differentiator based on photonics", *IEEE Transactions on Circuits and Systems: Express Briefs*, vol. 67, no.11, pp. 2767-2771, 2020. DOI:10.1109/TCSII.2020.2965158.

49. M. Tan, et al., "Photonic RF arbitrary waveform generator based on a soliton crystal micro-comb source", Journal of Lightwave Technology, vol. 38, no. 22, pp. 6221-6226 (2020). DOI: 10.1109/JLT.2020.3009655.

50. M. Tan, X. Xu, J. Wu, R. Morandotti, A. Mitchell, and D. J. Moss, "RF and microwave high bandwidth signal processing based on Kerr Micro-combs", Advances in Physics X, VOL. 6, NO. 1, 1838946 (2021). DOI:10.1080/23746149.2020.1838946.

51. X. Xu, et al., "Advanced RF and microwave functions based on an integrated optical frequency comb source", Opt. Express, vol. 26 (3) 2569 (2018).

52. M. Tan, X.Xu, J. Wu, B. Corcoran, A. Boes, T. G. Nguyen, S. T. Chu, B. E. Little, R.Morandotti, A. Lowery, A. Mitchell, and D. J. Moss, ""Highly Versatile Broadband RF Photonic Fractional Hilbert Transformer Based on a Kerr Soliton Crystal Microcomb", Journal of Lightwave Technology 39 (24) 7581-7587 (2021).

53. Wu, J. et al RF Photonics: An Optical Microcombs' Perspective. IEEE Journal of Selected Topics in Quantum Electronics Vol. **24**, 6101020, 1-20 (2018).

54. T. G. Nguyen et al., "Integrated frequency comb source-based Hilbert transformer for wideband microwave photonic phase analysis," *Opt. Express*, vol. 23, no. 17, pp. 22087-22097, Aug. 2015.

55. X. Xu, et al., "Continuously tunable orthogonally polarized RF optical single sideband generator based on micro-ring resonators," *Journal of Optics*, vol. 20, no. 11, 115701. 2018.

56. X. Xu, et al., "Orthogonally polarized RF optical single sideband generation and dual-channel equalization based on an integrated microring resonator," *Journal of Lightwave Technology*, vol. 36, no. 20, pp. 4808-4818. 2018.

57. X. Xu, et al., "Photonic RF phase-encoded signal generation with a microcomb source", *J. Lightwave Technology*, vol. 38, no. 7, 1722-1727, 2020.

58. X. Xu, et al., Broadband microwave frequency conversion based on an integrated optical micro-comb source", *Journal of Lightwave Technology*, vol. 38 no. 2, pp. 332-338, 2020.

59. M. Tan, et al., "Photonic RF and microwave filters based on 49GHz and 200GHz Kerr microcombs", *Optics Comm.* vol. 465,125563, Feb. 22. 2020.

60. M. Tan et al, "Orthogonally polarized Photonic Radio Frequency single sideband generation with integrated micro-ring resonators", IOP Journal of Semiconductors. Vol. **42** (4), 041305 (2021). DOI: 10.1088/1674-4926/42/4/041305.

61. Mengxi Tan, X. Xu, J. Wu, T. G. Nguyen, S. T. Chu, B. E. Little, R. Morandotti, A. Mitchell, and David J. Moss, "Photonic Radio Frequency Channelizers based on Kerr Optical Micro-combs", IOP Journal of Semiconductors Vol. **42** (4), 041302 (2021). DOI:10.1088/1674-4926/42/4/041302.

62. B. Corcoran, et al., "Ultra-dense optical data transmission over standard fiber with a single chip source", Nature Communications, vol. 11, Article:2568, 2020.

63. X. Xu et al, "Photonic perceptron based on a Kerr microcomb for scalable high speed optical neural networks", Laser and Photonics Reviews, vol. 14, no. 8, 2000070 (2020). DOI: 10.1002/lpor.202000070.

64. X. Xu, et al, "11 TOPs photonic convolutional accelerator for optical neural networks", Nature **589**, 44-51 (2021).

65. Xingyuan Xu, Weiwei Han, Mengxi Tan, Yang Sun, Yang Li, Jiayang Wu, Roberto Morandotti, Arnan Mitchell, Kun Xu, and David J. Moss, "Neuromorphic computing based on wavelength-division multiplexing", *IEEE Journal of Selected Topics in Quantum*

*Electronics* **29** (2) 7400112 (2023). DOI:10.1109/JSTQE.2022.3203159.

66. Yang Sun, Jiayang Wu, Mengxi Tan, Xingyuan Xu, Yang Li, Roberto Morandotti, Arnan Mitchell, and David Moss, "Applications of optical micro-combs", Advances in Optics and Photonics **15** (1) 86-175 (2023). DOI:10.1364/AOP.470264.

67. Yunping Bai, Xingyuan Xu,1, Mengxi Tan, Yang Sun, Yang Li, Jiayang Wu, Roberto Morandotti, Arnan Mitchell, Kun Xu, and David J. Moss, "Photonic multiplexing techniques for neuromorphic computing", Nanophotonics **12** (5): 795–817 (2023). DOI:10.1515/nanoph-2022-0485.

68. Chawaphon Prayoonyong, Andreas Boes, Xingyuan Xu, Mengxi Tan, Sai T. Chu, Brent E. Little, Roberto Morandotti, Arnan Mitchell, David J. Moss, and Bill Corcoran, "Frequency comb distillation for optical superchannel transmission", Journal of Lightwave Technology **39** (22) 7383-7392 (2021). DOI: 10.1109/JLT.2021.3116614.

69. Mengxi Tan, Xingyuan Xu, Jiayang Wu, Bill Corcoran, Andreas Boes, Thach G. Nguyen, Sai T. Chu, Brent E. Little, Roberto Morandotti, Arnan Mitchell, and David J. Moss, "Integral order photonic RF signal processors based on a soliton crystal micro-comb source", IOP Journal of Optics **23** (11) 125701 (2021). https://doi.org/10.1088/2040-8986/ac2eab

70. Yang Sun, Jiayang Wu, Yang Li, Xingyuan Xu, Guanghui Ren, Mengxi Tan, Sai Tak Chu, Brent E. Little, Roberto Morandotti, Arnan Mitchell, and David J. Moss, "Optimizing the performance of microcomb based microwave photonic transversal signal processors", accepted Sept. 1, *Journal of Lightwave Technology* **41** (23) pp 7223-7237 (2023), doi: 10.1109/JLT.2023.3314526.

71. Mengxi Tan, Xingyuan Xu, Andreas Boes, Bill Corcoran, Thach G. Nguyen, Sai T. Chu, Brent E. Little, Roberto Morandotti, Jiayang Wu, Arnan Mitchell, and David J. Moss, "Photonic signal processor for real-time video image processing based on a Kerr microcomb", *Communications Engineering* **2** 94 (2023). DOI: 10.1038/s44172-023-00135-7.

72. Mengxi Tan, Xingyuan Xu, Jiayang Wu, Roberto Morandotti, Arnan Mitchell, and David J. Moss, "Photonic RF and microwave filters based on 49GHz and 200GHz Kerr microcombs", Optics Communications, 465, 125563 (2020). doi:10.1016/j.optcom.2020.125563. doi.org/10.1063/1.5136270.

73. Yang Sun, Jiayang Wu, Yang Li, Mengxi Tan, Xingyuan Xu, Sai Chu, Brent Little, Roberto Morandotti, Arnan Mitchell, and David J. Moss, "Quantifying the Accuracy of Microcomb-based Photonic RF Transversal Signal Processors", *IEEE Journal of Selected Topics in Quantum Electronics* **29** no. 6, pp. 1-17, Art no. 7500317 (2023). 10.1109/JSTQE.2023.3266276.

74. Yang Li, Yang Sun, Jiayang Wu, Guanghui Ren, Xingyuan Xu, Mengxi Tan, Sai Chu, Brent Little, Roberto Morandotti, Arnan Mitchell, and David Moss, "Feedback control in micro-comb-based microwave photonic transversal filter systems", Special Issue on Optical Microcombs, *IEEE Journal of Selected Topics in Quantum Electronics* (IF=5) (2024). DOI: 10.1109/JSTQE.2024.3377249.

75. Weiwei Han, Zhihui Liu, Yifu Xu, Mengxi Tan, Yuhua Li, Xiaotian Zhu, Yanni Ou, Feifei Yin, Roberto Morandotti, Brent E. Little, Sai Tak Chu, Xingyuan Xu, David J. Moss, and Kun Xu, "Dual-polarization RF Channelizer Based on Microcombs", *Optics Express* **32**, No. 7, 11281 (2024). DOI: 10.1364/OE.519235.

76. Aadhi A. Rahim, Imtiaz Alamgir, Luigi Di Lauro, Bennet Fischer, Nicolas Perron, Pavel Dmitriev, Celine Mazoukh, Piotr Roztocki, Cristina Rimoldi, Mario Chemnitz, Armaghan Eshaghi, Evgeny A. Viktorov, Anton V. Kovalev, Brent E. Little, Sai T. Chu, David J. Moss, and Roberto Morandotti, "Mode-locked laser with multiple timescales in a microresonator-based nested cavity", *APL Photonics* **9** 031302 (2024); (2024). 10.1063/5.0174697.

77. C. Mazoukh, L. Di Lauro, I. Alamgir1 B. Fischer, A. Aadhi, A. Eshaghi, B. E. Little, S. T. Chu, D. J. Moss, and R. Morandotti, "Genetic algorithm-enhanced microcomb state generation", *Communications Physics* Vol. 7, Article: 81 (2024) (2024). DOI: 10.1038/s42005-024-01558-0.

78. Yang Li, Yang Sun, Jiayang Wu, Guanghui Ren, Bill Corcoran, Xingyuan Xu, Sai T. Chu, Brent. E. Little, Roberto Morandotti, Arnan Mitchell, and David J. Moss, "Processing accuracy of microcomb-based microwave photonic signal processors for different input signal waveforms", *MDPI Photonics* **10**, 10111283 (2023). DOI:10.3390/photonics10111283.

79. Yang Sun, Jiayang Wu, Yang Li, and David J. Moss, "Comparison of microcomb-based RF photonic transversal signal processors implemented with discrete components versus integrated chips", *MDPI Micromachines* **14**, 1794 (2023). DOI: 10.3390/mi14091794.

80. Yang Sun, Jiayang Wu, Yang Li, Xingyuan Xu, Guanghui Ren, Mengxi Tan, Sai Tak Chu, Brent E. Little, Roberto Morandotti, Arnan Mitchell and David J. Moss, "Optimizing the performance of microcomb based microwave photonic transversal signal processors", *Journal of Lightwave Technology* **41** (23) 7223-7237 (2023). DOI: 10.1109/JLT.2023.3314526.

81. Yang Sun, Jiayang Wu, Yang Li, Mengxi Tan, Xingyuan Xu, Sai Chu, Brent Little, Roberto Morandotti, Arnan Mitchell, and David J. Moss, "Quantifying the Accuracy of Microcomb-based Photonic RF Transversal Signal Processors", *IEEE Journal of Selected Topics in Quantum Electronics* vol. 29, no. 6, 7500317 (2023). DOI: 10.1109/JSTQE.2023.3266276.

82. Bao, C., et al., Direct soliton generation in microresonators, Opt. Lett., **42**, 2519 (2017).